\documentclass[floatfix,aip,graphicx,preprint,apl]{revtex4-1}

\usepackage{graphicx}% Include figure files
\usepackage{dcolumn}% Align table columns on decimal point
\usepackage{bm}% bold math
\usepackage[utf8]{inputenc}
\usepackage{chemformula} % Formula subscripts using \ch{}
\usepackage[T1]{fontenc} % Use modern font encodings
\usepackage[euler]{textgreek}
\usepackage{float}
\usepackage[section]{placeins}
\usepackage{flafter}

\usepackage{listings}
\usepackage{xcolor}

\definecolor{mygreen}{rgb}{0.0, 0.6, 0.0}
\definecolor{mygray}{rgb}{0.5, 0.5, 0.5}
\definecolor{mymauve}{rgb}{0.58, 0.0, 0.82}
\definecolor{backcolour}{rgb}{0.95,0.95,0.92}

\usepackage{mathptmx}
\usepackage{etoolbox}

\usepackage{amsmath}
\usepackage{amssymb}
\usepackage{xcolor}

\DeclareUnicodeCharacter{2032}{\ensuremath{^{\prime}}}

\newcommand{\etal}{\mbox{\textit{et al.}\:}} %make et al. italic 

\makeatletter
\def\@email#1#2{%
 \endgroup
 \patchcmd{\titleblock@produce}
  {\frontmatter@RRAPformat}
  {\frontmatter@RRAPformat{\produce@RRAP{*#1\href{mailto:#2}{#2}}}\frontmatter@RRAPformat}
  {}{}
}%
\makeatother

\begin{document}

%\preprint{AIP/123-QED}

\title[A step-by-step workflow to account for linear polarization artifacts in circular dichroism of thin films]{A step-by-step workflow to account for linear polarization artifacts in circular dichroism of thin films}

\author{Franziska Schölzel}
%\email{salvan@physik.tu-chemnitz.de}
  \affiliation{Institute of Physics, Chemnitz University of Technology, Reichenhainer Str. 70, 09126 Chemnitz, Germany}
 \affiliation{Research Center for Materials, Architectures and Integration of Nanomembranes, Chemnitz University of Technology (MAIN), Chemnitz}
%\thanks{}

\author{Arina Narudin}
 \affiliation{Institute of Physics, Chemnitz University of Technology, Reichenhainer Str. 70, 09126 Chemnitz, Germany}

\author{Aleksandra Ciesielska}
 \affiliation{Hybrid Materials Design (HyMaD), Institute for Materials Research (IUMAT),
Hasselt University, Martelarenlaan 42, Hasselt B-3500, Belgium}
 \affiliation{EnergyVille, IUMAT, Thor Park 8320, 3600 Genk, Belgium} 
 \affiliation{imec, IUMAT, Wetenschapspark 1, B-3590, Diepenbeek, Belgium}

\author{Alexander Ehm}
 \affiliation{Institute of Physics, Chemnitz University of Technology, Reichenhainer Str. 70, 09126 Chemnitz, Germany}

\author{Dietrich R.T. Zahn}
 \affiliation{Institute of Physics, Chemnitz University of Technology, Reichenhainer Str. 70, 09126 Chemnitz, Germany}
 \affiliation{Research Center for Materials, Architectures and Integration of Nanomembranes, Chemnitz University of Technology (MAIN), Chemnitz}

\author{Wouter Van Gompel}
 \affiliation{Hybrid Materials Design (HyMaD), Institute for Materials Research (IUMAT),
Hasselt University, Martelarenlaan 42, Hasselt B-3500, Belgium}
\affiliation{EnergyVille, IUMAT, Thor Park 8320, 3600 Genk, Belgium}
\affiliation{imec, IUMAT, Wetenschapspark 1, B-3590, Diepenbeek, Belgium}
 
\author{Simon Kahmann$^*$}
 \email{simon.kahmann@physik.tu-chemnitz.de}
 \affiliation{Institute of Physics, Chemnitz University of Technology, Reichenhainer Str. 70, 09126 Chemnitz, Germany}

\author{Georgeta Salvan$^*$}
  \affiliation{Institute of Physics, Chemnitz University of Technology, Reichenhainer Str. 70, 09126 Chemnitz, Germany}
 \affiliation{Research Center for Materials, Architectures and Integration of Nanomembranes, Chemnitz University of Technology (MAIN), Chemnitz}
 \email{salvan@physik.tu-chemnitz.de}

\date{\today}

\begin{abstract}
Circular Dichroism (CD) spectroscopy has evolved from a purely solution based method towards an important tool in the analysis of chiral thin films. Although a straightforward technique, the true CD signal is often accompanied by artifacts arising from optical anisotropy and instrumental imperfections. Here, we present a two-step workflow that separates the orientation invariant CD response for anisotropic thin films by combining azimuthal sample rotation with sample flipping. For this purpose, a home-built sample stage was developed, which enables systematic suppression of many anisotropy-induced artifacts in commercial CD spectrophotometers. Both a detailed description of the setup itself as well as the required python script are provided. The reliability of the workflow is demonstrated on two selected samples from different research fields: chiral molecules attached to metallic surfaces as well as low-dimensional metal halide perovskites incorporating chiral spacer molecules.
\end{abstract}


\maketitle %\maketitle must follow title, authors, abstract and \pacs

%%%%%%%%%%%%%%%%%%%%%%%%%%%%%%%%%%%%%%%%%%%%%%%%%%%%%%%%%%%%%%%%%%%%%

\section{Introduction}
Chirality or handedness describes the existence of two variations - enantiomers - of a substance that are mirror opposites of each other \cite{moss1996basic}. Nowadays, chirality plays a central role in a wide range of research fields, spanning from biomolecules \cite{ranjbar2009circular} over organic semiconductors \cite{yang2017emergent} to halide perovskites \cite{wang2025structural} and chiral superstructures \cite{albano2020chiroptical,lv_self-assembled_2022}. Distinguishing between the two enantiomers is often a challenging task. Currently, the most powerful way to differentiate enantiomers is based on their interaction with light. Chiral materials typically exhibit chiroptical activity, which modifies the polarization state of light. This can be probed by circular dichroism (CD) or optical rotary dispersion (ORD) spectroscopy. CD and ORD effects occur when left-handed (LCP) and right-handed circular polarized (RCP) light is absorbed and/or retarded to different extents by a chiral material. \\
As the research on chiral materials has its roots in biochemistry, CD has been widely used for the analysis of enantiomers in solution. Driven in part by the discovery of the chirality induced spin selectivity (CISS) effect, interest in chiral structures has also expanded towards thin film systems and their applications in fields such as spintronics, electrochemistry and magnetic resonance\cite{li2025spatial, chiesa2021assessing, wei2021chiral}. This led to unexpected challenges in the field of CD spectroscopy. Most importantly, chiral molecules in solution are randomly oriented, whereas in the solid state they can adopt a preferential alignment. Such reduced degree of freedom related to preferential molecular alignment can induce uniaxial or biaxial optical anisotropy of the films \cite{troxell1971electric}, giving rise to linear dirchroism (LD) and/or linear birefringence (LB) in addition to CD \cite{troxell1971electric, kuroda2001solid, Salij_Apparent_CD}. The disentanglement of the CD, LD, and LB contributions from the measured signal in CD spectroscopy is therefore essential, yet challenging, and is often neglected. For example, in the field of chiral perovskite thin films, the measured CD spectrum can reverse sign when the sample surface is flipped by 180\,$^\circ$ relative to the light propagation direction, a behaviour that would not be expected if the signal originated from CD only\cite{albano2020chiroptical,Salij_Apparent_CD,zhang2022revealing}. Although contributions from LD and LB to the measured CD spectra have already been reported in an earlier review article\cite{kuroda2001solid}, and have received more attention in recent publications \cite{albano2020chiroptical, zhang2022revealing}, a simple and broadly accessible experimental workflow for isolating reliable CD signals in thin films is still lacking. Accurate accounting of LD and LB artifacts is complicated because the required sample stage and electronics are not provided as standard equipment by commercial vendors of CD spectrometers.\\
In this work, we highlight the fundamental effects that can obscure the true CD response of solid samples exhibiting uniaxial or biaxial optical anisotropy. By presenting a two-step workflow for extracting a linear artifact-corrected CD signal for anisotropic thin film samples, we aim to make robust thin film CD measurements more accessible. The workflow is based on a home-built sample stage that enables systematic suppression of anisotropy induced artifacts in CD measurements with commercial CD spectrophotometers. The feasibility of the two-step method is demonstrated using two selected examples from timely research fields: chiral molecules attached to metallic surfaces and metal halide perovskites incorporating chiral spacer molecules. Together, these examples illustrate potential pitfalls in measuring CD spectra of thin films while underscoring the necessity of the two-step workflow for obtaining robust and reproducible results.\\

\section{Background on circular dichroism spectroscopy and measurement artifacts}
\begin{figure}[tbh]
    \centering
    \includegraphics[width=0.9\linewidth]{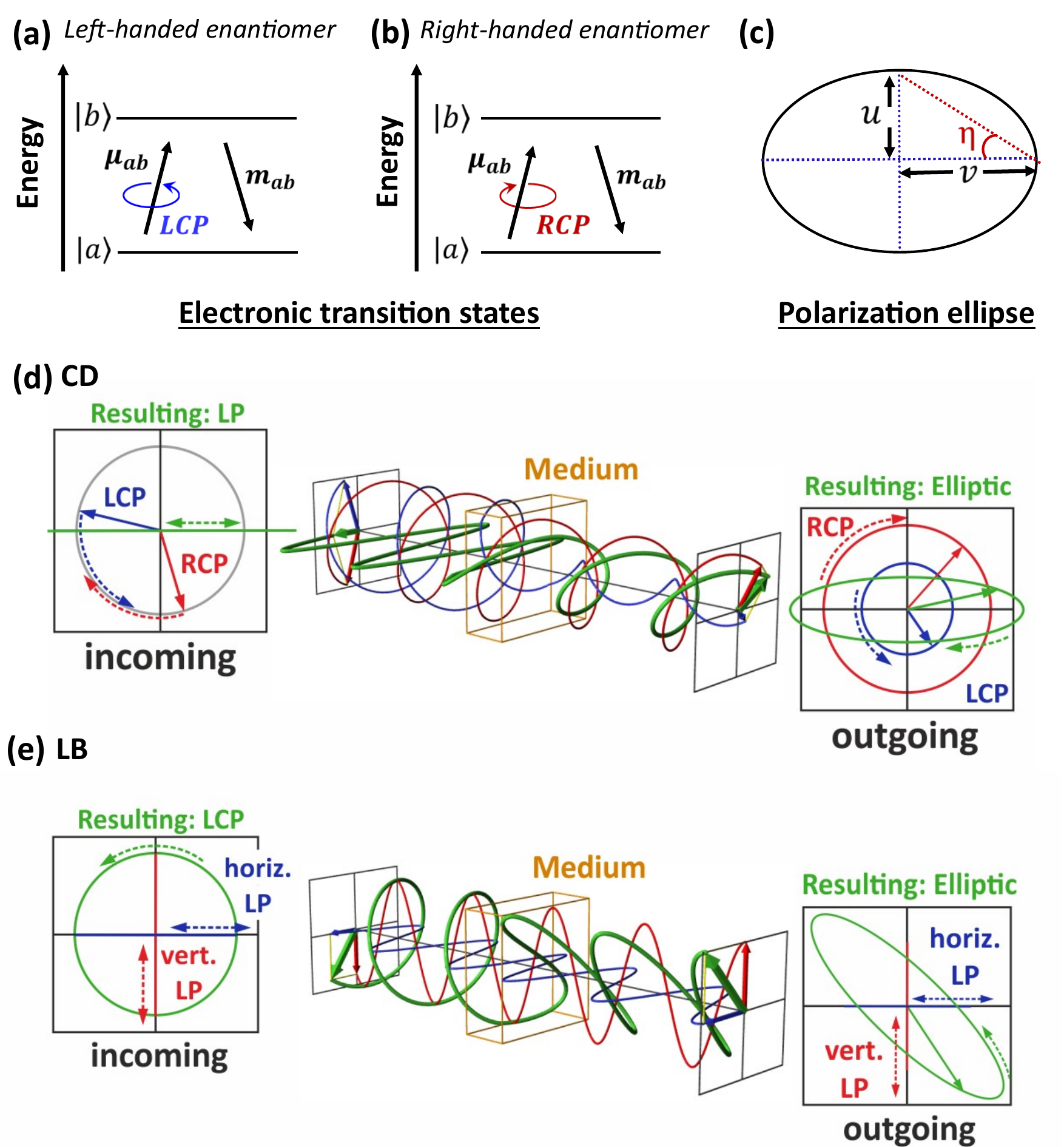}
    \caption{(a, b) Energy-level diagrams illustrating the microscopic origin of CD. Optical activity arises from the simultaneous electric ($\mu_\text{ab}$) and magnetic ($m_\text{ab}$) transition dipole moments associated with an electronic transition between states $\langle a|$ and $\langle b|$ of two enantiomers of the same chiral substance, leading to preferential interaction with (a) LCP or (b) RCP. (c) Polarization ellipse characterised by principal axes, $u$ and $v$, and the ellipticity angle, $\eta$. (d) CD: An incident LP light (green) can be described as superposition of two in-phase components of LCP (blue) and RCP (red) of equal amplitude. Preferential absorption of either results in elliptically polarized light (green). (e) LB: anisotropic refractive indices introduce a phase delay between orthogonal linearly polarization components, converting a circular polarized wave into an elliptically polarized output. [(d)-(e) were generated using: EMANIM \cite{EManiMwebsite}].}
    \label{Fig. Polarisation}
\end{figure}
\subsection{How thin films affect the polarization of light}

To understand the working mechanism of our two-step measurement routine, we first introduce both the microscopic and the macroscopic descriptions of CD, along with additional macroscopic effects that can influence the CD measurements. 
CD arises from the differential absorption of LCP and RCP light in chiral media, i.e., media lacking mirror symmetry. Microscopically speaking, it results from differences in the transition probabilities for left- and right-circularly polarized photons during a transition from a ground state, $a$, to an excited state, $b$. This phenomenon was formulated by Léon Rosenfeld \cite{rosenfeld1929quantenmechanische} and is illustrated in Figs.~\ref{Fig. Polarisation}~(a) and (b). The magnitude of this effect is proportional to the rotatory strength $R (a \rightarrow b)$, defined as the imaginary part of the scalar product between the electric dipole transition moment $\langle a| \boldsymbol{\mu}|b\rangle$ and the magnetic dipole transition moment $\langle b| \mathbf{m}|a\rangle$:\cite{berova2012comprehensive}
\begin{equation}
\\R (a \rightarrow b) \sim \operatorname{Im}(\langle a \mid \boldsymbol{\mu} \mid b\rangle \cdot \langle b \mid \mathbf{m} \mid a \rangle)
\label{dipoles}
\end{equation}
In essence, CD requires that an electric dipole transition is accompanied by a magnetic dipole contribution with a non-zero relative phase. In practical terms, this leads to different absorption probabilities of circularly polarized $\sigma ^+$ (RCP) and $\sigma ^-$ (LCP) photons. For example, consider a chiral molecule with a single electronic transition, such as an amino acid. For the L-enantiomer, the transition probability is higher for $\sigma ^-$ -polarized light, resulting in more $\sigma ^-$ photons than $\sigma ^+$ photons being absorbed \cite{Kobayashi2011CDOrgChem} (See Fig.~S3 (a)). Of course, this is reversed for the D-enantiomer, where the transition probability is higher for $\sigma ^+$. In more complex systems such as chiral perovskites, the interaction with circularly polarized light (CPL) depends not only on the handedness of the material but also on the nature of the underlying electronic transition, resulting in both $\sigma ^+$ and $\sigma ^-$ dominated transitions for the same enantiomer but at different energies. Thus, one cannot directly predict the chirality of a substance by the preferred handedness of light. Nevertheless, differential absorption provides a direct experimental means for distinguishing the enantiomers of a chiral substance.
Experimentally, CD is quantified by measuring the transmitted intensities of RCP and LCP light at a given wavelength and evaluating the corresponding difference in absorbance\cite{berova2012comprehensive}: 
\begin{equation}
\Delta A_\text{CD} = A_\text{LCP} - A_\text{RCP}
\label{CD}
\end{equation}
This differential absorbance directly reflects the microscopic rotatory strength of the underlying electronic transitions. In practice, CD is often reported in terms of ellipticity $\eta$, which is related to differential absorption by: 
\begin{equation}
\eta~(\mathrm{mdeg}) = 32.98 \times \Delta A_{\mathrm{CD}}
\label{Ellipticity}
\end{equation}
A detailed derivation of Eq.~\ref{Ellipticity} and the rationale for expressing CD in terms of ellipticity are provided in the Supplementary Note II. 
Because light exhibits both particle and wave characteristics, CD can also be described within the framework of classical electromagnetic theory. This macroscopic, wave-based description provides a more straightforward and intuitive treatment and is adopted throughout the remainder of this text. Within this framework, the superposition of two circularly polarized components, RCP and LCP waves, is represented as a linearly polarized (LP) plane electromagnetic wave (see the incoming wave in Fig.~\ref{Fig. Polarisation} (d)). Upon propagation through a chiral medium, these circular components retain their polarization states, making them the normal modes of the medium. If the medium absorbs RCP and LCP waves to a different extent, their superposition after propagation yields an elliptical polarization state, as shown in Fig.~\ref{Fig. Polarisation} (d). This effect corresponds to CD and is commonly quantified via ellipticity \cite{berova2012comprehensive}:
\begin{equation}
\eta\,(\text{mdeg}) = tan^{-1}\biggl(\frac{u}{v}\biggr)
\end{equation}
whereby $u$ and $v$ denote the minor and major axes of the polarization ellipse (corresponding to the projection of the electric field vector on the ellipse axes; see Fig.~\ref{Fig. Polarisation} (c)). This explains why CD is not unitless, but is instead expressed in millidegrees (mdeg). 

If, instead, the refractive indices for the RCP and LCP waves differ, the two modes propagate at different velocities, resulting in a relative phase shift. Their superposition remains linearly polarized, but with a rotated polarization direction relative to the original wave. This phenomenon, known as circular birefringence (CB) or ORD (Fig.~S1 (a)), lays the historical foundation of spectroscopic CD measurements. It can be readily observed in optically isotropic media, either liquids or solids. The wavelength (or photon energy) dependence of both CD and CB is often used to gain insight into the electronic structure of chiral materials.

The above mentioned changes in the polarization state (elliptization and rotation) of light, however, are not caused solely by chiral materials. In optically anisotropic but achiral materials, the two normal modes are LP waves, and can produce similar effects. These "measurement artifacts" are in fact intrinsic optical properties of the anisotropic samples. Linear dichroism (LD), Fig.~S1 (b), describes the light-matter interaction of anisotropic materials that absorb two orthogonal LP waves to a different extent \cite{berova2012comprehensive, rodger2016linear}, leading to a tilt of the polarization direction with respect to the original direction, i.e., similar to the CB effect. In linear birefringence (LB) (Fig.~\ref{Fig. Polarisation} (e)) the two LP normal modes experience different velocities in the material and therefore, accumulate a phase delay \cite{shrivastava2018introduction}. Their superposition leads to an elliptical polarization state that mimics the CD effect. In addition, when LD and LB occur simultaneously in an oriented sample, combined LDLB effects arise, where the material both differentially absorbs the normal light modes and introduces a phase shift between them.\\ 
More generally, absorption and dispersion are fundamentally linked: the spectral behaviour of these effects is linked through the Kramers-Kronig relations\cite{ferre1984linear}. Therefore, CD, CB, LD, and LB may co-exist in materials that are both chiral and anisotropic, requiring careful experimental procedures to isolate the intrinsic ("clean") CD response.

\subsection{Artifacts in CD measurements}

To date, three experimental approaches have been employed to measure CD. Early measurements, such as those on heme proteins, used modified double-beam spectrophotometers to directly detect absorption differences. However, these differences in absorption are too small to be reliably detected for many systems\cite{berova2012comprehensive}. The second approach measures the ellipticity induced in linearly polarized light (LPL) passing through the sample, which is limited by the difficulty of maintaining LPL quality and detecting the small orthogonal components of elliptically polarized light (EPL). Third, modern CD spectrometers most commonly use rapid polarization modulation, typically with a photoelastic modulator (PEM), which significantly enhances sensitivity, signal-to-noise ratio, and enables reliable detection of weak CD signals\cite{berova2012comprehensive}. 

Although PEM-based techniques are highly sensitive, the measured CD signal can still include contributions from the optical anisotropy of the film or substrate rather than from intrinsic sample chirality, resulting in spurious CD. CD artifacts in thin film samples arise from three main sources - some of which can be disentangled, whereas others cannot.

The first principal source of artifacts is LDLB in the sample, which during polarization modulation can produce an apparent intensity difference between LCP and RCP, generating a false CD signal comparable in magnitude to the intrinsic signal.
A second source of artifacts arises from instrumental imperfections. Modern CD spectrometers use a PEM to rapidly switch between LCP and RCP and synchronously detect the differential signal. The PEM induces periodic birefringence in a transparent crystal, converting LPL into alternating circularly and elliptically polarized states at a modulation frequency $\sim$ $50$\,kHz \cite{berova2012comprehensive}. Ideally, this enables isolation of the true CD signals, but in practice, residual linear polarization components can be superimposed with the sample response, thereby producing spurious CD-like signals.\\ 
A third source of artifacts arises from wavelength-dependent scattering, reflections, or thin-film interference within the sample or substrate, all of which can alter the polarization state and produce signals unrelated to molecular chirality \cite{castiglioni2010evaluation}. Unlike LDLB, such effects are difficult to disentangle, as they depend on factors including surface roughness, film thickness, refractive-index contrast, and measurement geometry, and may yield spectral features that mimic intrinsic CD signals; thus, making them virtually impossible to remove \cite{ugras2023can}. 
Therefore, careful experimental design, appropriate control measurements, and rigorous data analysis are essential to reliably disentangle the intrinsic CD from artifacts. 

\subsection{Understanding and mitigating linear contributions in CD: A Stokes–Mueller perspective}

To quantitatively disentangle the linear and circular contributions in CD measurements, it is useful to describe the optical response of the sample within the Stokes–Mueller formalism, which provides a general framework for analyzing the interaction between light and anisotropic media. In this formalism, the polarization state of light is represented by a four-component Stokes vector that encodes the total intensity as well as the degree of linear and circular polarization. The effect of an optical element or sample on this polarization state is then described by a Mueller matrix, which linearly transforms the incident Stokes vector into the output Stokes vector. This approach explicitly accounts for how linear optical properties can influence the detected circular signal through polarization modulation. 

Within this framework, the experimentally measured CD signal does not necessarily correspond solely to true CD. Instead, it can contain additional contributions arising from LD, LB, and coupling terms between these effects. These coupling mechanisms originate from the fact that the polarization state of the probing light evolves as it propagates through anisotropic media, allowing linear optical anisotropies to convert or mix polarization components that are detected as apparent circular signals. \\
Previous studies have applied this formalism to mitigate linear artifacts, addressing macroscopic anisotropy contributions, identify LDLB coupling terms, and integrate microscopic LDLB modelling within Lorentz oscillator models \cite{kuroda2001solid, kuroda2024fast, ugras2023can, salij2021theory}.
The Mueller matrix is a $4 \times 4$ matrix that describes how an optical systems transforms the Stokes vector of incident light, thereby providing a complete phenomenological description of linear, circular, and depolarizing effects. In principle, the measured CD signals can be quantitatively modelled through matrix calculations (Supplementary Note III). 

In practice, however, anisotropic thin films present additional challenges. Depolarization, scattering, and linear anisotropies are all embedded in the full Mueller matrix, and extracting solid information requires measurement of multiple matrix elements. Partial measurements are generally insufficient because no single Mueller matrix element can be uniquely assigned to CD when LDLB are intrinsically coupled \cite{takechi2011chiroptical}. This limitation is exacerbated in thin film systems whereby they are compounded by structural anisotropy, surface reflections, and scattering, that can introduce additional polarization distortions that can complicate the interpretation of the measured signal. 

Although full Mueller polarimetry can, in principle, disentangle these contributions by simultaneously quantifying birefringence, diattenuation, depolarization, and optical activity, such measurements become essential only when a rigorous quantitative decomposition of the polarization response is required \cite{arwin2021optical}. Consequently, despite rigorous framework provided by the Stokes-Mueller formalism, its full experimental implementation is often impractical for routine measurements of anisotropic thin films because of the associated instrumental complexity, measurement time, and data analysis requirements. 
To address this limitation, we adopt a more experimentally accessible strategy based on a two-step workflow. Specifically, we combine front/back acquisition with rotational CD measurements to reduce and diagnose linear anisotropy contributions in the measured spectra. These complementary measurements do not require full Mueller polarimetry, providing a practical approach for disentangling corrected CD signals.

\section{Measurement workflow}

\begin{figure}[htb]
    \centering
    \includegraphics[width=\linewidth]{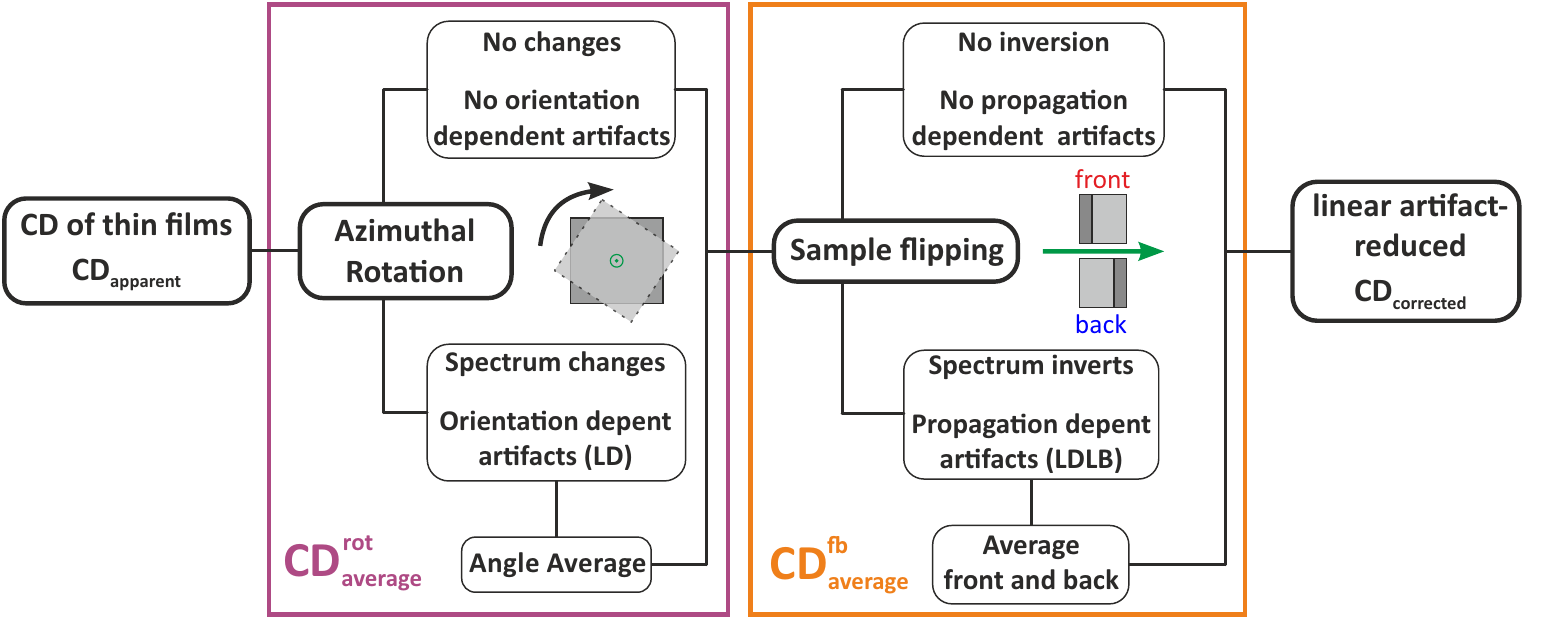}
    \caption{Overview of the two-step workflow for diagnosing and reducing artifacts in CD measurements. The first step involves azimuthal rotation which probes orientation-dependent effects ($CD^\text{rot}_\text{average}$), followed by sample flipping to identify propagation-dependent LDLB contributions ($CD^\text{fb}_\text{average}$). The recommended protocol for extracting corrected CD is summarized below.}
    \label{Fig. Workflow}
\end{figure}

In this workflow, depicted in Fig.~\ref{Fig. Workflow}, we present a two-step protocol that incorporates azimuthal rotation and flipping. Both methods are essential for robust measurements, as they remove distinct classes of artifacts.  Experimentally, the measured CD response can be expressed as: \cite{zhang2022revealing, makhija2024effect, albano2017chiroptical}
\begin{equation}
CD_{\text{apparent}} \approx CD_{\text{corrected}} + \frac{1}{2} \left( LD' \, LB - LB' \, LD \right) + (- LD) \, \sin\alpha \, \cos 2\theta
\label{Eq. CD apparent} %please don't change labels
\end{equation}
Here, the first term corresponds to the intrinsic CD contribution. The second term, $\frac{1}{2}(LD^\prime LB - LB^\prime LD)$, is a rotation-invariant apparent CD arising from coupling between LD and LB. In this expression, $LD^\prime$ and $LB^{\prime}$ denote the $LD$ and $LB$ measured along axes rotated by $45\,^\circ$ relative to $0\,^\circ/90\,^\circ$axes, which is necessary because the anisotropy in thin films is not generally aligned with the measurement frame. The third term, $(-LD)\, \sin\alpha \, \cos 2\theta$, accounts for a rotation-dependent artifact from residual birefringence in the PEM $\alpha$ \cite{zhang2022revealing}, $\theta$ is the azimuthal angle, while the negative sign indicates that this contribution has an opposite sign relative to the LD defined along the $0\,^\circ/90\,^\circ$ axes. 
From Eq.~\ref{Eq. CD apparent} it becomes clear that azimuthal rotation suppresses contributions arising from in-plane optical anisotropy that depend on the sample orientation (term 3), whereas flipping eliminates linear effects that reverse sign under propagation inversion (term 2). Only the combination of both enables us to obtain the corrected CD signal in thin films; neither rotation nor flipping alone is sufficient.

\textbf{Step 1 — Azimuthal rotation; Purpose: identify anisotropy contributions and remove orientation dependence.}
The first strategy we employ in this work is the angle-dependent CD measurement, which is widely regarded as a robust approach, particularly for highly anisotropic samples \cite{narushima2016circular, zhang2022revealing, makhija2024effect}. In this step, the thin film is rotated about the propagation axis of the incident light by an azimuthal angle, $\theta$, while the CD spectrum ($CD^\text{rot}(\theta)$) is recorded. The purpose of this step is diagnostic: in anisotropic thin films, the measured CD can contain a rotation-dependent contribution that varies as $\cos2\theta$, arising from LD in the film  ($LD \neq 0$) combined with imperfect circular polarization of the PEM. Observing such angular modulation provides insight that linear artifacts are contributing to the apparent CD signal.\\
In this case, if the CD signal changes with $\cos2\theta$, especially in regions of large absorption, that is a strong indication of linear artifacts, not intrinsic chirality. This step allows us to diagnose anisotropy contributions before disentangling them. 
To further substantiate that the observed rotation-dependent CD signals originate from linear artifacts rather than intrinsic chirality, we show in Supplementary Note V, a Mueller matrix analysis of an achiral reference sample based on the work of Kuroda~\etal which explicitly demonstrates the LB-coupling artifacts \cite{kuroda2001solid}. 

After collecting $CD^\text{rot}(\theta)$ at uniformly spaced azimuthal angles over a range spanning from  $0^\circ$ to $180^\circ$ (with $+15\,^\circ$ increments) or better $0^\circ$ to $360^\circ$ (with $+30\,^\circ$ increments), to collect one full period of the expected cosine dependence, the spectra are averaged to remove orientation-dependent contributions. The chosen angular spacing is sufficient to capture the rotational dependence while maintaining a practical measurement time. With $n$ denoting the total number of measured azimuthal angles, this yields:
\begin{equation}
    CD_\text{average}^\text{rot}\approx\frac{\sum_{i=0}^{n} CD^\text{rot}(\theta)}{n}
\end{equation}
This procedure eliminates the $\cos2\theta$-dependent term because it integrates to zero over a full angular cycle. As a result, rotation-dependent artifacts are effectively decoupled. However, rotation-invariant contributions, including true CD and LDLB, may still remain at this stage. A step by step description of how the rotational averaging procedure works and its effect in eliminating artifacts is provided in the Supplementary Note IV. 

\textbf{Step 2 — Sample flipping; Purpose: suppress artifacts linked to propagation direction and extraction of corrected CD.} 
Next, we employ the sample flipping protocol, originally introduced by Shindo~\etal and later formalized by Kuroda~\etal \cite{kuroda2001solid, zhang2022revealing}. In this approach, the CD spectrum is first measured with incident light passing through the film, yielding the front-side spectrum ($CD_\text{front}$), and the spectrum is recorded again with light entering from the opposite direction (the substrate), yielding the back-side spectrum ($CD_\text{back}$). Because genuine CD is a pseudoscalar, it remains invariant upon flipping, whereas linear effects reverse sign as they depend on the optical axis orientation. 

Ideally, this step corrects for all linear artifacts that cancel under propagation inversion. Consequently, any component of the signal that changes sign upon flipping is generally interpreted as evidence of linear contamination. This argument, however, strictly holds only for normal transmission geometry and may break down in the presence of reflection contributions, optical path asymmetries, or interference effects between multiple layers or interfaces, all of which can modify the measured signal in a direction-dependent manner.  
Nevertheless, the influence of these artifacts can be reduced by averaging the front and back CD spectra, which leads to the cancellation of the sign-changing linear contributions. The inversion-independent CD, denoted as $CD_\text{average}^\text{fb}$, is thus obtained as  \cite{zhang2022revealing}:
\begin{equation}
CD_\text{average}^\text{fb} \approx \frac{1}{2} (CD_{\text{front}} + CD_{\text{back}}) 
\label{Eq. CD genuine}
\end{equation}
Here, the approximation sign accounts for additional minor artifacts, such as scattering, reflection, and interference. Conversely, if we take the difference between the two measurements, it will provide us a direct estimation of the linear contributions \cite{makhija2024effect, zhang2022revealing}. 
\begin{equation}
LDLB^\text{fb} \approx \frac{1}{2} (CD_{\text{front}} - CD_{\text{back}}) 
\label{Eq. LDLB}
\end{equation}
Therefore, by sequentially applying the entire two-step protocol: azimuthal rotation and sample flipping, we progressively reduce rotation-dependent and propagation-direction-dependent artifacts. We combine these steps to:
\begin{equation}
    CD_\text{corrected}\approx\frac{1}{2}\biggl(\frac{\sum_{i=0}^{n} CD^\text{rot}_\text{front}(\theta)+\sum_{i=0}^{n} CD^\text{rot}_\text{back}(\theta)}{n}\biggr)
    \label{Eq. CD genuine average}
\end{equation}
for an extraction of the corrected CD response. And for the contribution of the linear artifacts to:
\begin{equation}
    LDLB\approx\frac{1}{2}\biggl(\frac{\sum_{i=0}^{n} CD^\text{rot}_\text{front}(\theta)-\sum_{i=0}^{n} CD^\text{rot}_\text{back}(\theta)}{n}\biggr)
    \label{Eq. LDLB average}
\end{equation}

\textbf{Quantitative evaluation of the chiroptical activity.} Having established a practical workflow for extracting the linear artifact-corrected CD, we next turn to the quantitative evaluation of the CD signal strength and, therefore, chiroptical activity. To allow a direct comparison between samples of different thicknesses, it is convenient to introduce the figure of merit, $g_\text{CD}$, which is defined as \cite{zhang2022revealing, makhija2024effect, ishii2020direct}: 
\begin{equation}
    g_\text{CD}=\frac{CD~[\text{mdeg}]}{32\,980 \times Absorbance~[\text{o.d.]}}
    \label{Eq. g_CD}
\end{equation}
The $g_\text{CD}$ value normalizes the measured CD signal to the absorbance of the chiral material at a certain wavelength. When applying this metric to thin film measurements, one should also perform the above correction protocol to the absorbance spectra to account for substrate absorption/CD. Additionally, the substrate must be measured and the according spectra subtracted from the films as otherwise, the resulting $g_\text{CD}$ value would correspond to the layer stack rather than to the chiroptical material itself.

Overall, these considerations underscore that, despite the widespread availability and sophistication of modern CD instrumentation, the rigorous discrimination between intrinsic chiroptical response and contributions arising from linear optical anisotropies remains a non-trivial experimental and analytical challenge. This issue is particularly pronounced in anisotropic thin film systems, where even subtle linear optical artifacts can be partially converted into apparent CD, thereby complicating quantitative interpretation and potentially leading to significant overestimation or misassignment of the true CD signal.

\section{Experimental implementation} % Franziska
\begin{figure}[tbh]
\centering
    \includegraphics[width=\linewidth]{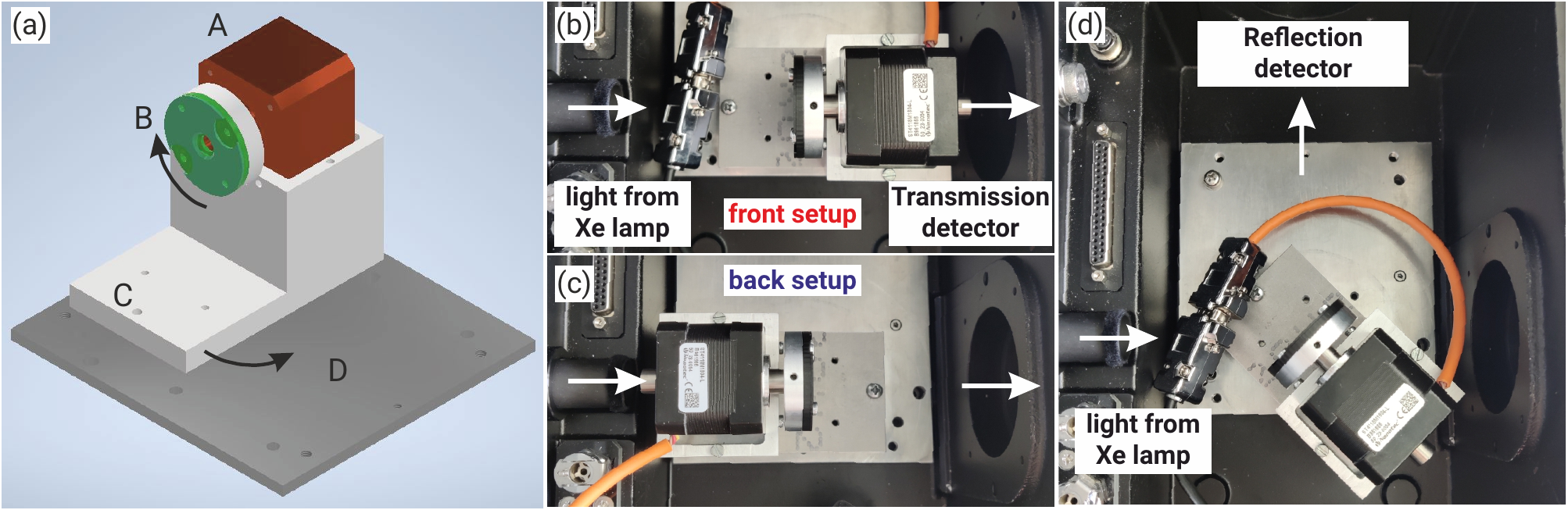}
    \caption{(a) Technical drawing of the designed rotation mount with the components (A) step motor, (B) sample stage, (C) fixating screw, (D) baseplate. The arrows indicate the rotation directions. Picture of sample compartment of CD spectrometer with built-in rotation mount in (b) front, (c) back and (d) reflection configuration.}
    \label{fig. Rotation Mount}
\end{figure}

In the following, we present our custom-built CD extension for a commercial CD spectrometer aiming to reduce linear artifacts and provide a step-by-step description of the measurement process.

Our CD measurements were performed on a J-1500 spectrophotometer from JASCO, for which we designed an automated rotation mount as depicted in Fig.~\ref{fig. Rotation Mount}~(a). It consists of a baseplate (Fig.~\ref{fig. Rotation Mount}~(a) D) that can be screwed to the bottom of the sample compartment of the J-1500 to prevent movement. A step motor mount located on top of the base plate can be rotated in 45$\,^\circ$ steps as indicated by the black arrow. This allows the mount to be positioned reproducibly in front and back orientation, depicted in Figs.~\ref{fig. Rotation Mount}~(b) and (c), as well as at an angle of 45\,$^\circ$ to enable reflection measurements (see Fig.~\ref{fig. Rotation Mount}~(d)). The step motor mount and the base plate are connected with a dowel pin and the position is fixed again with a screw marked C in Fig.~\ref{fig. Rotation Mount}~(a). The motor is a NEMA 17 stepper with a hollow shaft from Nanotec and has a precision of 0.02$\,^\circ$. Importantly, the hollow shaft allows light transmission. The motor is controlled by a 2-phase SBC-MD-DM860H driver from Joy-it run by a python script, which is provided in the Supplementary Information. This allows the user to freely choose both the step size and the maximum azimuthal angle. The sample is placed on a sample stage, marked as B in Fig.~\ref{fig. Rotation Mount}~(a), consisting of an aluminum base plate with a 3D printed sample holder on top. The sample space can be varied depending on the sample dimensions ranging from $(5\times5)$\,mm$^2$ to $(25\times25)$\,mm$^2$. The sample is fixed with a cover plate.

To enable automated azimuthal rotation and measurements, the motor controlling program and the measurement software are synchronized with the help of the \textit{Macro Command} Program from JASCO. Both programs must run simultaneously. The workflow of the resulting measurement is as follows: First, the user sets the CD measurement parameters in the Jasco \textit{Macro Command}. Afterwards a spectrum is measured. As soon as the measurement file and the exported file are saved, the stage is rotated. To accommodate for the rotation time, a delay in the form of a Jasco "wait box" is added between measurements. This scheme can be continued until the desired maximum angle is reached. The \textit{Macro Command} Program Code is supplied in Fig.~S5, followed by the Python code that drives the step motor in Listing~1 in the SI.

\section{Experimental validation of the workflow}
To demonstrate the functionality of the setup described above, we investigated how optical anisotropy influences the CD signal across two different sample systems. First, we build the bridge to the origin of CD spectroscopy by considering chiral molecular thin films, while also exploring the applicability of thin metal films on sapphire as substrates for transmission CD measurements. In addition, we also studied the CD response of metal halide perovskites incorporating chiral spacer molecules.

\subsection{Example 1 - molecular assemblies on metal surfaces}

\begin{figure}[tbh]
    \centering
    \includegraphics[width=0.95\linewidth]{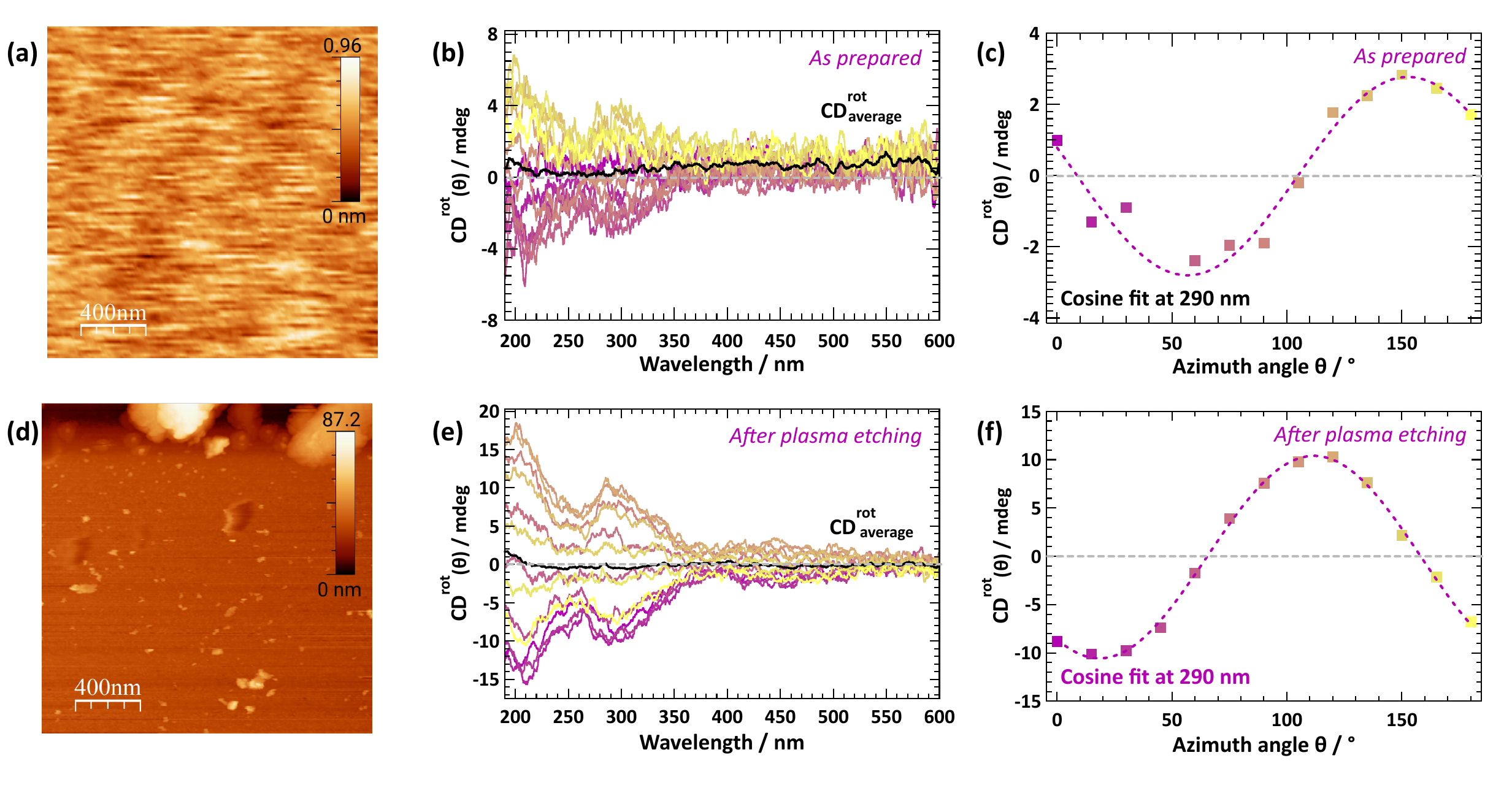}
    \caption{(a)/(d) $(1.3\times 1.3)\,\text{\textmu m}^2$ non-contact AFM images, (b)/(e) measured $CD^\text{rot}(\theta)$ spectra in front configuration for different azimuthal angles, the black line corresponding to $CD_\text{average}^\text{rot}$ (c)/(f) $CD^\text{rot}(\theta)$ values taken at 290\,nm in dependence of the azimuth angle with a cosine fit; for a 30\,nm Au film on sapphire (a)/(b)/(c) as prepared and (d)/(e)/(f) after plasma etching.}
    \label{fig. CD AFM Au}
\end{figure}
\noindent
We artificially roughened a 30\,nm Au film deposited on a c-plane sapphire with Argon ions via plasma etching (PE) to create an inhomogeneous sample surface. Due to its low thickness, it was possible to measure the film in transmission configuration. As Au forms achiral conformations, the sample should not exhibit any CD. However, due to reflections/scattering on surface structures, a CD signal is detected as depicted in Fig.~\ref{fig. CD AFM Au} (b). For simplicity, we only show the front measurements, the complementary plots for the back can be found in Fig.~S7 in the Supplementary Information. The different coloured lines correspond to individual measurements ($CD^\text{rot}(\theta)$) with a difference in azimuthal angle $\theta$ of 15\,$^\circ$. A CD signal is observed that varies with the azimuthal orientation of the sample, although the surface of the as-prepared Au film has a roughness below 1\,nm (Fig.~\ref{fig. CD AFM Au} (a)). To further evaluate the effect, the roughness of the film was increased by PE with Argon for five minutes. This increases the roughness by an order of magnitude as deduced from the AFM images (Fig.~\ref{fig. CD AFM Au} (d)) (also see height histograms and line profiles in Fig.~S6). As a result, the CD signal doubles in intensity (Fig.~\ref{fig. CD AFM Au} (e)). For both roughnesses, the optical anisotropy is clearly visible via the dependence of the CD signal on the azimuthal rotation of the sample, displayed in Figs.~\ref{fig. CD AFM Au} (c) and (f), by taking a closer look at the variation of $CD^\text{rot}(\theta)$ at one wavelength (290\,nm). Here, a clear cosine dependence of the signal on twice the azimuthal angle $\theta$ is observed. This correlation can be linked to Eq.~\ref{Eq. CD apparent} as the LD component has the same relation with $\theta$. When averaging over all measured spectra, it becomes clear that the signal is no true CD signal as it vanishes completely as indicated by the mean displayed as black lines in Figs.~\ref{fig. CD AFM Au} (b) and (e). This is true for both roughnesses as well as the front and the backside measurements (see Fig.~S7). To further underline the importance of sample rotation, we additionally extracted the front and back average ($CD^\text{fb}_\text{average}$) with Eq.~\ref{Eq. CD genuine} for all measured angles. Here, in stark contrast to the theory stating that the average between front and back measurements is able to eliminate artifacts, again a cosine dependence can be observed (cf. Fig.~S8). This proves that the front/back approach can only cancel certain types of artifacts but cannot account for reflection/scattering at surfaces, as these often depend on the propagation direction of the light. In consequence, these results demonstrate the necessity of averaging over different azimuthal angles to reduce artifacts emerging from reflection/scattering processes.
\begin{figure}[htb]
    \centering
    \includegraphics[width=0.9\linewidth]{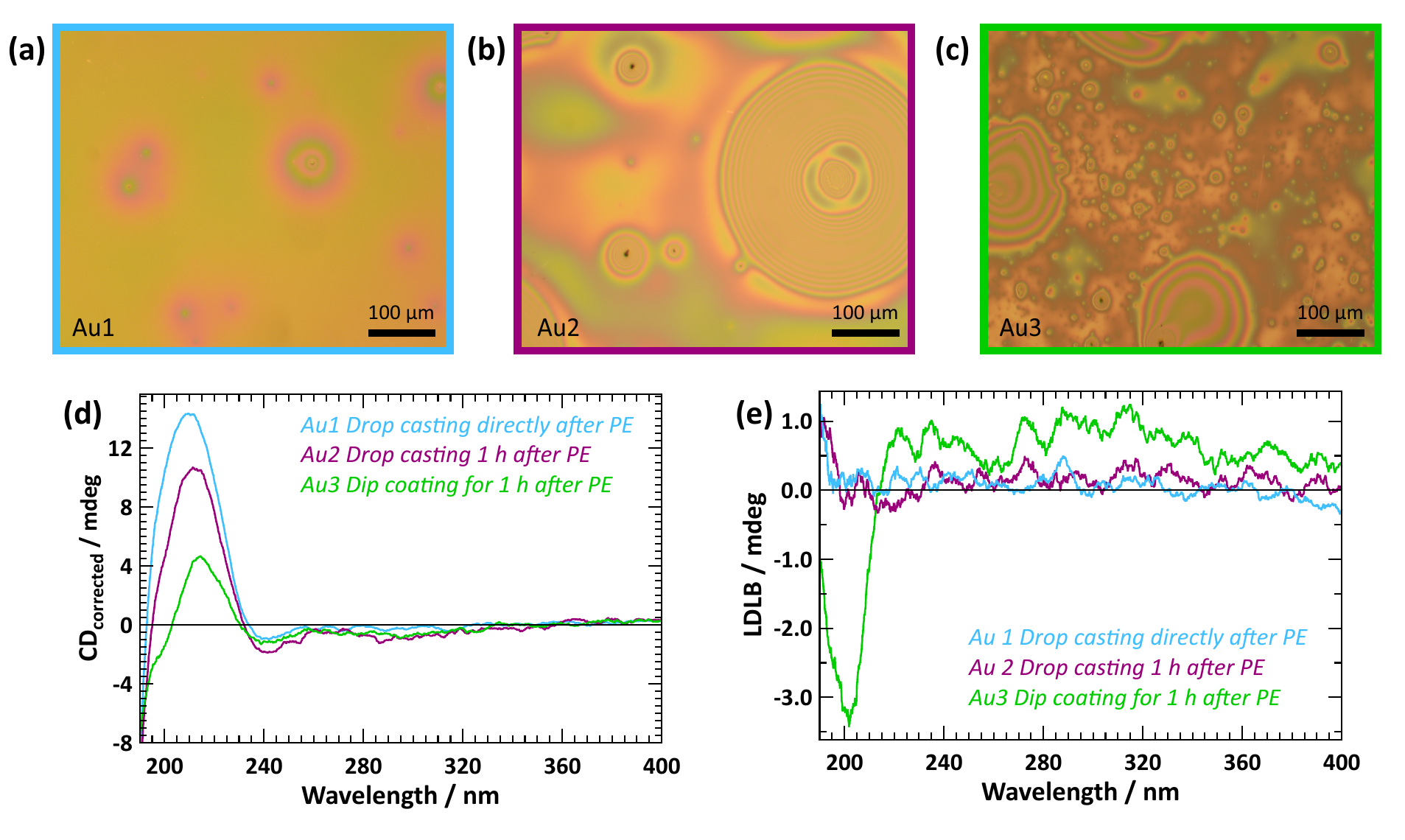}
    \caption{Optical Microscopy images with 20$\times$ magnification of Boc-L-Cysteine deposited via (a) drop coating directly after the PE process (Au1), (b) drop coating 1\,h after PE (Au2) and (c) dip coating for 1\,h (Au3); (d) extracted $CD_\text{corrected}$ and (e) $LDLB$ with for the samples shown in (a)-(c).}
    \label{fig. Cysteine LDLB}
\end{figure}
\noindent

If averaging over azimuthal rotation can reduce artifacts arising from surface reflection/scattering what can the front and back approach account for? To answer this question, we analysed the CD response and extracted the LDLB contribution after Eq.~\ref{Eq. LDLB average} of the amino acid Boc-L-Cysteine on Au. When comparing the $CD_\text{corrected}$ spectra, calculated with Eq.~\ref{Eq. CD genuine average}, of three different films prepared with different deposition techniques ranging from drop casting (Au1 and Au2) to dip coating (Au3) (see Supplementary Information for more details) a decrease in the CD intensity corresponding to the signal of Boc-L-Cysteine at 210\,nm is observed. As the signal intensity is directly correlated to the number of molecules, which are hit by the light beam, this difference can be explained by a decrease in film thickness from Au1 to Au3\cite{Kobayashi2011CDOrgChem}. This claim is supported by the layer thickness determined from variable-angle spectroscopic ellipsometry (VASE) measurements with $(460\pm30)$\,nm, $(340 \pm 20)$\,nm, and $(190 \pm 10)$\,nm for Au1 – Au3, respectively. Additionally, a change in lineshape between the first two samples and Au3 can be observed. This is mirrored in the LDLB spectra displayed in Fig.~\ref{fig. Cysteine LDLB} (e) by a peak arising visibly above noise level for Au3 at around 200\,nm displaying a difference between the front and back measurements. By averaging over these two one can reduce the influence of the LDLB artifact in the CD spectrum. For the other samples only noise around zero could be detected in the LDLB emphasizing the absence of orientation dependent artifacts.

As already mentioned in section III, for the quantitative evaluation of the chiroptical activity, the value of $g_\text{CD}$ is usually calculated. Here, a comparable to or even higher than the one of molecules on solution or comparable chiral substances can be a good indicator for a trustable CD response, as the $g_\text{CD}$ is an intrinsic value of the material itself. We note that the extracted $g_\text{CD}$ of the film contains information on the molecular packing, orientation, excitonic coupling, or interaction with the substrate, which might lead to either subtle or pronounced differences with respect to the $g_\text{CD}$ of the solution, and can be exploited for the characterization of the solid-state effects in molecular films. Other factors influencing the $g_\text{CD}$ might be scattering and optical interference effects. The scattering effects are minimized in our approach by the azimuthal rotation, while the optical interference effects are minimized by the careful alignment of the light wavevector parallel to the sample normal. With this in mind, we calculated the $g_\text{CD}$ with Eq.~\ref{Eq. g_CD} for the Boc-L-Cysteine films on Au. Although all three show a clean $CD_\text{corrected}$ spectrum the $g_\text{CD}$ reveals remaining anisotropies that could not be accounted for. We calculated the values at 210\,nm, the position of the amino acid signal from the $CD_\text{corrected}$ spectra. From the solution measurements of Boc-Cysteine in Ethanol (Fig.~S9) we could determine the $g_\text{CD}$ to be $(3.5\pm0.4)\times 10^{-3}$. For both drop cast films the values are comparable with $(4.2\pm0.4)\times 10^{-3}$ for Au1 and $(3.4\pm0.3)\times 10^{-3}$ for Au2 (for error estimation please refer to section VIII in the Supplementary Information). Both samples exhibit close to no LDLB contribution in Fig.~\ref{fig. Cysteine LDLB} (e). In comparison, the dip coated film has a $g_\text{CD}$ of $(0.9\pm0.4)\times 10^{-3}$, which is one order of magnitude too small. To find an explanation one has to look closer into what $g_\text{CD}$ contains. As the CD spectrometer measures in transmission it cannot distinguish between absorbed and reflected light resulting in an increase in absorbance if the surface has many scattering centers. This is true for sample Au3 as it shows a grainy structure with small crystals in Fig.~\ref{fig. Cysteine LDLB} (c). From this one can draw the conclusion that the incoming light is reflected on the structures leading to a higher absorbance and smaller $g_\text{CD}$. Besides this, a high depolarization was observed with VASE for the samples Au2 and Au3, which partially reached 100\% as depicted in Figs.~S12 (a)-(c). This is typically caused by a high degree of surface roughness or thickness non-uniformity. Interestingly, although both samples Au2 and Au3 are very inhomogeneous it only shows in the $g_\text{CD}$ value of the dip coated film (Au3). We therefore suspect that Au2 is on the brink of anisotropy that is needed for strong changes in the $g_\text{CD}$ value as well as a measurable LDLB contribution.\\
These results demonstrate the importance of the front and back approach. We have shown how it unmasks artifacts from orientation anisotropies of the film extracting the LDLB component while simultaneously extracting corrected CD spectrum. Additionally, this is a handy approach to estimate the film quality of a sample as strongly deviating $g_\text{CD}$ values between solution and film can hint towards strong anisotropy or artifacts our approach can not account for. In cases where the calculated LDLB signal differs from zero (such as sample Au3), additional independent investigations of the optical anisotropy by e.g. optical polarization microscopy, imaging ellipsometry, full Muller-matrix polarimetry, or reflection-mode analysis under azimuthal rotation might deliver valuable information on the microcrystalline structure of the molecular films.

\subsection{Example 2 - chiral halide perovskite thin films}

Chiral halide perovskites and perovskite-inspired structures\cite{kahmann2025pathways} have emerged as a new paradigm for the design of chiral functional materials, owing to their tunable chemical composition and hierarchical structures\cite{wang2025structural}. Their outstanding properties render them promising candidates for applications in optoelectronic devices. Moreover, the perovskite framework can accommodate a wide variety of chiral organic ligands. By selecting appropriate enantiomeric organic cations, one can direct the formation of diverse chiral metal-halide structures; including 0D clusters, 1D chains, 2D and quasi-2D layered structures, as well as 3D networks\cite{kwon2025ligand}.

\begin{figure}[htb]
    \centering
    \includegraphics[width=1.0\linewidth]{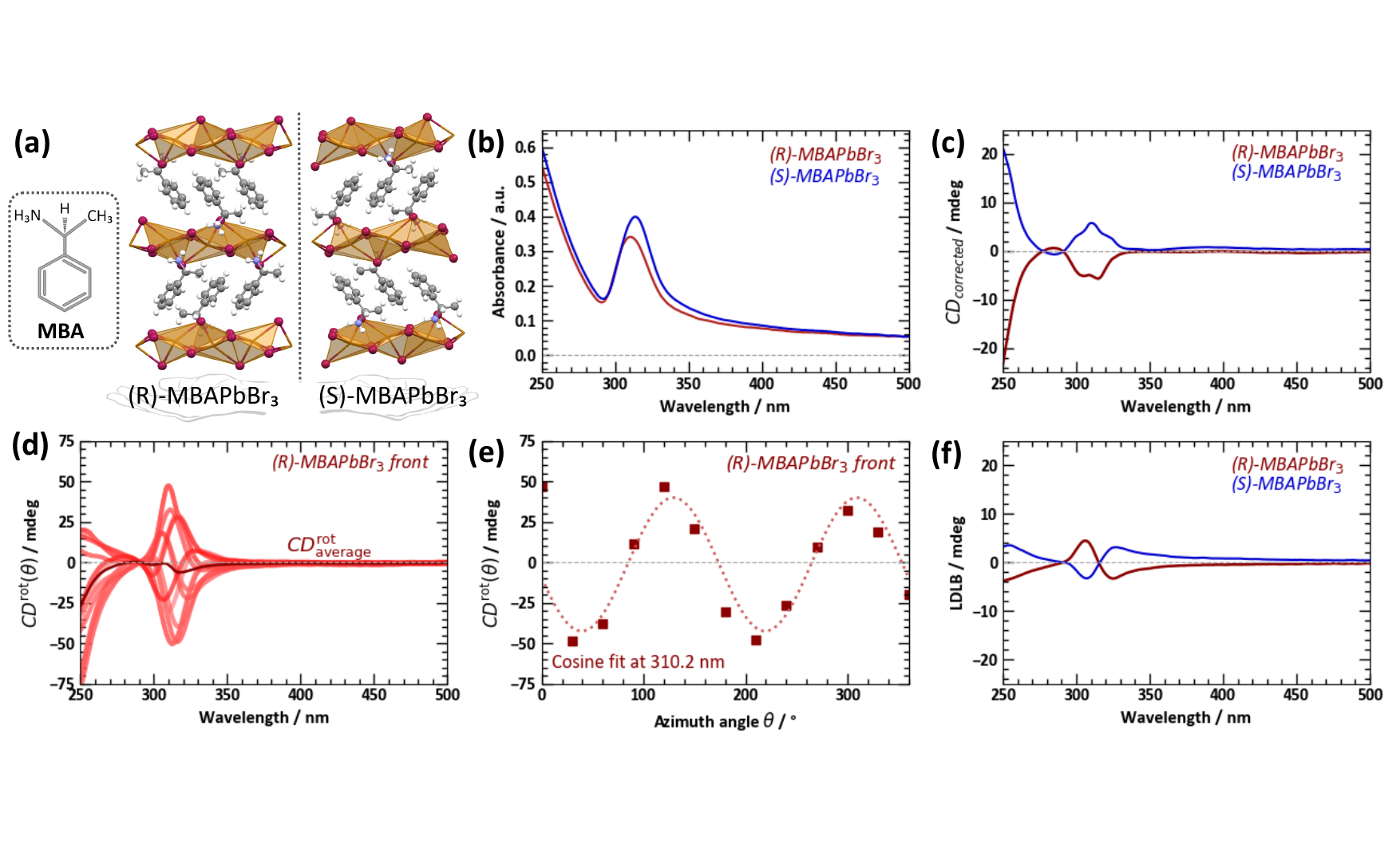}
    \caption{1D chiral perovskites thin films. (a) Crystal structures of (R)- and (S)-MBAPbBr$_3$, featuring the chiral MBA organic spacers. (b) Absorption spectra of (R)- and (S)-MBAPbBr$_3$ thin films measured from the front side. (c) $CD_\text{corrected}$ spectra of (R)- and (S)-MBAPbBr$_3$ . (d) Angle-dependent CD spectra of the (R)-MBAPbBr$_3$ film acquired from the front side of the film at different azimuthal angles. The dark red line in between is the $CD^\text{rot}_\text{average}$. (e) CD as a function of azimuthal angle at the excitonic peak; the red squares denote the experimental CD values, and the red dashed line shows a cosine-function fit. (f) LDLB contributions of both 1D films.}
    \label{Fig:pero}
\end{figure}

In this study, thin films of a one-dimensional (1D) and two-dimensional (2D) chiral halide perovskite were fabricated using the prototypical methylbenzylammonium (MBA\textsuperscript{+}) cations in its two enantiomeric forms: (R)-(+)-MBA\textsuperscript{+} and (S)-(–)-MBA\textsuperscript{+}. The use of enantiopure organic cations allows the transfer of molecular chirality to the extended inorganic framework.\\
The 1D compound, (R/S)-MBAPbBr$_3$, adopts the space group P2$_1$2$_1$2$_1$, consisting of face-sharing [PbBr$_6$]$^{4-}$ octahedra forming an infinite chains oriented along the [002] direction\cite{makhija2024effect, dang2020bulk, lu2021spin}. These inorganic chains are spatially separated by intercalated chiral MBA\textsuperscript{+} cations, as illustrated in Fig.~\ref{Fig:pero}~(a). Each octahedron is structurally asymmetric, exhibiting six distinct Pb–Br bond lengths and lacking both inversion and mirror symmetry, features that contribute directly to the material’s chiral nature\cite{billing2006synthesis, lu2021spin, makhija2024effect}.
In comparison, the 2D perovskites, (R/S)-MBA$_2$PbI$_4$, adopt a layered structure composed of a corner-sharing [PbI$_6$]$^{4-}$octahedra network, with each octahedron exhibiting six distinct Pb–I bond lengths, and with the layers separated by chiral MBA\textsuperscript{+} cations, as shown in Fig.~S16~(a).

Because MBA\textsuperscript{+} cations can template both 1D and 2D perovskite phases depending on stoichiometry and processing conditions, we ensured the formation of the correct phase using  X-ray diffraction (XRD) measurements for both systems. For the 1D phase, by comparison with a simulated powder XRD pattern\cite{dang2020bulk}, we find the peaks corresponding to the (002) and (004) reflections of the 1D phase located at $8.63\,^{\circ}$ and $17.31\,^{\circ}$, respectively (See Fig.~S18). These results are in agreement with those previously reported by Lu~\etal \cite{lu2021spin} and with our observation of an excitonic peak around $\sim$320\,nm in the absorption spectra, characteristic of the 1D bromide perovskite framework (Fig.~\ref{Fig:pero} (b)). \\  
Meanwhile, for 2D (R)- and (S)-MBA$_2$PbI$_4$, the XRD pattern of the thin films depicted in Fig.~S19 matches the one reported by Soares~\etal, confirming the formation of the targeted 2D phase structure \cite{soares2025chirality}. In addition, the excitonic peak located around $\sim$498\,nm further corroborates the characteristic signature of 2D iodide perovskites. 

Perovskite thin films are typically polycrystalline and frequently exhibit varying degrees of preferred orientation. As above, structural anisotropy can give rise to significant LD and LB that corrupt the measured CD signal\cite{zhang2022revealing}. To disentangle these linear optical artifacts from the measured CD signal,
we first determine the angle-dependent CD measurements over a full $0\,^{\circ}$ to $360\,^{\circ}$ rotation with +$30\,^{\circ}$ increments. Both 1D (R)- and (S)-MBAPbBr$_3$ films exhibit CD signals that vary systematically with rotation angle, displaying a clear cosine dependence at their respective excitonic peaks (310.2\,nm for (R)-MBAPbBr$_3$ and 313.9\,nm for (S)-MBAPbBr$_3$; Figs.~\ref{Fig:pero} (e) and S15 (b)). 
Next, we apply our second strategy by acquiring CD spectra from both front and back sides for all the measured angles of (R)- and (S)-MBAPbBr$_3$ and applying Eq.~\ref{Eq. CD genuine}; the associated LDLB terms were then calculated using Eq.~\ref{Eq. LDLB}. 
As shown in Fig.~S13 (b), 1D films display substantial $LDLB^\text{fb}$ contributions. Notably, the CD front- and back-side spectra of both systems do not match as expected (see Figs.~S14 (b) and ~S15 (a)), indicating strong optical anisotropy. In such cases, taking the front-back average cannot fully eliminate the linear artifacts, as averaging alone does not fully cancel these orientation-dependent effects\cite{salij2021theory}. 
Following the workflow, performing the angle-dependent CD measurements and averaging the CD spectra over all rotation angles diminish the propagation-dependent LDLB effects in the chiral perovskites (See Fig.~\ref{Fig:pero} (f)) and yield a reliable estimation of the corrected CD \cite{makhija2024effect, zhang2022revealing}.

In contrast, 2D systems exhibit small or negligible linear contributions ($LDLB^\text{fb}$) as illustrated in Fig.~S16 (d). For these samples, the flipping method is, in principle, sufficient to obtain the corrected CD. Nevertheless, we still need to assess whether any residual artifacts remain, such as linear optical effects or scattering-related contributions, by performing azimuthal rotation measurements. Upon rotation, there are no changes in the CD spectra and no cosine dependence is observed (see Figs.~S17 (a, b) and (c)), indicating that rotation-dependent artifacts are negligible in these samples. Minor scattering contributions may still be present, but they appear to be insignificant. 

Because 1D (R)- and (S)-MBAPbBr$_3$ films exhibit substantial LDLB contributions, we further quantified their chiroptical activity and CD signal strength by evaluating the $g_\text{CD}$ for both films. 
The corrected CD and absorbance spectra within the excitonic region were used to calculate $g_\text{CD}$ according to Eq.~\ref{Eq. g_CD}. This analysis yielded $g_\text{CD}=(-2.0 \pm0.6) \times 10^\text{-3}$ and $g_\text{CD}=(3.4 \pm0.7) \times 10^\text{-3}$ for (R)-MBAPbBr$_3$ and (S)-MBAPbBr$_3$, respectively. The uncertainty was estimated as described in Supplementary section VIII. \\
Our measured $g_\text{CD}$ fall within the typical range reported for 1D chiral perovskites thin films ($10^\text{-3}$ to $10^\text{-2}$)  \cite{zhang2022revealing, kwon2025ligand}, well below the values approaching $10^{-1}$ reported for a few highly engineered 1D systems \cite{guo2025tunable, asad2024realization}. 
Based on this comparison, the $g_\text{CD}$ values obtained for (R)- and (S)-MBAPbBr$_3$, can be classified as lying in the lower-to-moderate regime \cite{kwon2025ligand, guo2025tunable}. The same procedure was then applied to the 2D perovskite systems, yielding $g_\text{CD}=(2.1 \pm0.1) \times10^\text{-4}$ and $g_\text{CD}=(-2.2 \pm0.1) \times 10^\text{-4}$ for (R)-MBA$_2$PbI$_4$ and (S)-MBA$_2$PbI$_4$, respectively. These values, and the associated uncertainties are notably smaller than those of 1D counterparts. Such reduced $g_\text{CD}$ values have been widely reported for MBA-based 2D perovskites \cite{makhija2024effect, kim2024giant}, where limited structural distortion and reduced long-range order hinder efficient chirality transfer to the inorganic framework \cite{dang2020bulk, moroni2024chiral}. 

\section{Limitation of the workflow}

Although the front/back flipping and azimuthal rotational CD workflow is effective in many cases, its applicability is fundamentally limited by two factors:

\textbf{Irreducible optical artifacts.}
During CD measurements, scattering, reflection, or other optical effects can occur from inhomogeneities in the thin film, such as surface roughness, grain boundaries, or thickness variations. These irregularities cause part of the incident light to deviate from the original propagation direction and often produce signals that differ between the front and back sides of the film. Such contributions have nothing to do with the CD and are considered as artifacts; they cannot be removed simply by flipping or rotation; they must be considered independently. \\
For example, our measurements on the gold film demonstrate that strong scattering at the surface due to high roughness produces an apparent CD signal that is purely artificial. Angle dependent CD measurements further revealed a cosine dependence, highlighting the presence of such artifacts. \\
Even under an ideal transmission geometry, flipping only removes linear effects that reverse sign under propagation inversion. It does not compensate for other direction-dependent optical contributions, such as scattering or reflections, meaning that flipping alone cannot guarantee that the corrected CD is free from all artifacts.

\textbf{Thin film anisotropy and optical symmetry.}
The workflow further assumes that the optical response of the sample is symmetric under propagation reversal. This assumption fails in anisotropic films. Consider 2D perovskites: although these materials are structurally anisotropic, the term “anisotropy” can be misleading in the context of CD measurements. CD is typically measured along the stacking (out-of-plane) direction, along which in-plane processes average out and appear isotropic. Consequently, the optical response probed in this geometry is largely insensitive to in-plane anisotropy, even though the material itself remains anisotropic at the microscopic level. \\
This distinction is critical: the workflow’s ability to disentangle linear artifacts depends on the optical anisotropy projected along the measurement direction, not on the intrinsic anisotropy of the material alone.

\section{Conclusion}

We have developed a robust two-step workflow for extracting linear artifact-corrected CD spectra from anisotropic chiral thin films. Central to this approach is a motorized sample stage that enables automated azimuthal rotation and straightforward front-back measurements. Schematic drawings and the corresponding source code are provided in the Supplementary Information. The reliability of the approach was validated by CD measurements on different samples.\\
Our first example highlights the importance of azimuthal rotation to eliminate artifacts arising from reflection or scattering processes at the surface of a thin Au film. Using the proposed workflow, the measured CD signals were identified as pure artifact-based. In addition, we showed the value of the front and back approach for extracting the corrected CD response of anisotropic films by varying the deposition technique of Boc-L-Cysteine on Au substrates. Furthermore, we underline how the LDLB component and the $g_\text{CD}$ value can be used as diagnostic tools for the estimation of the film quality. \\
Our workflow was further demonstrated and can also be applicable for chiral halide perovskite thin films, where 1D systems exhibit moderate $g_\text{CD}$ in the range of $10^\text{-3}$, confirming their chiroptical activity but readily susceptible to obscuration by linear artifacts, which can be partially mitigated using our approach. The remaining linear artifacts may arise from scattering or thin-film interference in 1D structures. In contrast, 2D systems show smaller $g_\text{CD}$ values ($10^\text{-4}$) and negligible LDLB contributions, with linear artifacts that can be disentangled using the workflow.
Overall, this work establishes practical guidelines and best practices for obtaining corrected CD from the linear artifacts in thin films. The presented approach is broadly applicable to chiral hybrid perovskites and other low-dimensional materials, enabling unambiguous interpretation of CD spectra and to avoid common pitfalls associated with orientation-dependent optical artifacts.\\

%%%%%%%%%%%%%%%%%%%%%%%%%%%%%%%%%%%%%%%%%%%%%%
\section*{Supplementary Material}
See supplementary material for additional information on the derivation of the equations used in the manuscript, detailed information about the Mueller matrix formalism, source codes for driving the developed rotation stage, and further results of the molecular films and the chiral halide perovskites.

\begin{acknowledgments}
F. S. and G. S. gratefully acknowledge funding through German Research Foundation (DFG), TRR 386, TP B3, project number 514664767. This project was co-financed by the European Union and by tax funds in accordance with the budget adopted by the Saxon state parliament (EFRE InfraProNet 2021-2027 NachLeuchten 100701886 and 100701890).
The authors thank Axel Fechner and Volker Krauß for their help with the construction of the rotation stage. Furthermore, F.S. and G.S. would like to thank the group of Prof. Thomas Seyller (TU Chemnitz) for the opportunity to use their AFM; and the Center of Materials (TU Chemnitz) for the deposition of the Au films.
A. C. acknowledges the Research Foundation - Flanders (FWO) for the funding of her FWO strategic basic research PhD grant (1SH0X24N). 
W.T.M.V.G. acknowledges the FWO for the funding of the WEAVE project INTENSITY (G0AQV25N).
A.E. and D.R.T.Z. gratefully acknowledge funding by the DFG Research Unit FOR 5387 POPULAR, project no. 461909888.
\end{acknowledgments}

\section{Author Declarations}
\textbf{Conflict of interest}\\
The authors declare no conflict of interest.

\section*{Author Contributions}

\textbf{Franziska Schölzel}: Writing -- original draft (equal); Writing -- review \& editing (equal); Methodology (Design and construction of the rotation stage; Measurements of molecular assemblies on metal surfaces); Data curation (equal). \textbf{Arina Narudin}: Writing -- original draft (equal); Writing -- review \& editing (equal); Methodology (Background on CD and measurement artifacts, Measurements on chiral perovskites, and limitations); Data curation (equal). \textbf{Aleksandra Ciesielska}: Writing -- review \& editing (supporting); Data curation (preparing chiral hybrid perovskite samples). \textbf{Alexander Ehm}: Writing -- review \& editing (supporting); Data curation (VASE measurements and fitting). \textbf{Dietrich R.T. Zahn}: Writing -- review \& editing (supporting); Supervision (supporting). \textbf{Wouter Van Gompel}: Writing -- review \& editing (supporting); Supervision (supporting); Methodology (chiral perovskite design \& synthesis). \textbf{Simon Kahmann}: Conceptualization (lead); Supervision (lead); Writing -- review \& editing (lead). \textbf{Georgeta Salvan}: Conceptualization (lead); Supervision (lead); Writing -- review \& editing (lead); Resources (Funding grant for the CD spectrometer)\\

\section*{Data Availability Statement}
The data that support the findings of this study and technical drawings of the rotation stage are available from the corresponding author upon reasonable request.

\bibliography{library}

@article{Salij_Apparent_CD,
author = {Salij, Andrew and Goldsmith, Randall H. and Tempelaar, Roel},
title = {Theory of Apparent Circular Dichroism Reveals the Origin of Inverted and Noninverted Chiroptical Response under Sample Flipping},
journal = {Journal of the American Chemical Society},
volume = {143},
number = {51},
pages = {21519-21531},
year = {2021}
}

@article{zhang2022revealing,
  title={Revealing the intrinsic chiroptical activity in chiral metal-halide semiconductors},
  author={Zhang, Zixuan and Wang, Zhiyu and Sung, Herman H-Y and Williams, Ian D and Yu, Zhi-Gang and Lu, Haipeng},
  journal={Journal of the American Chemical Society},
  volume={144},
  number={48},
  pages={22242--22250},
  year={2022},
  publisher={ACS Publications}
}

@article{albano2020chiroptical,
  title={Chiroptical properties in thin films of $\pi$-conjugated systems},
  author={Albano, Gianluigi and Pescitelli, Gennaro and Di Bari, Lorenzo},
  journal={Chemical reviews},
  volume={120},
  number={18},
  pages={10145--10243},
  year={2020},
  publisher={ACS Publications}
}

@article{kuroda2001solid,
  title={A solid-state dedicated circular dichroism spectrophotometer: Development and application},
  author={Kuroda, Reiko and Harada, Takunori and Shindo, Yohji},
  journal={Review of Scientific Instruments},
  volume={72},
  number={10},
  pages={3802--3810},
  year={2001},
  publisher={American Institute of Physics}
}

@article{billing2006synthesis,
  title={Synthesis and crystal structures of inorganic--organic hybrids incorporating an aromatic amine with a chiral functional group},
  author={Billing, David G and Lemmerer, Andreas},
  journal={CrystEngComm},
  volume={8},
  number={9},
  pages={686--695},
  year={2006},
  publisher={Royal Society of Chemistry}
}

@article{troxell1971electric,
  title={Electric dichroism and polymer conformation. I. Theory of optical properties of anisotropic media, and method of measurement},
  author={Troxell, Terry C and Scheraga, Harold A},
  journal={Macromolecules},
  volume={4},
  number={5},
  pages={519--527},
  year={1971},
  publisher={ACS Publications}
}

@article{lu2021spin,
  title={Spin-Dependent Charge Transport in 1D Chiral Hybrid Lead-Bromide Perovskite with High Stability},
  author={Lu, Ying and Wang, Qian and Chen, Ruyi and Qiao, Leilei and Zhou, Foxin and Yang, Xia and Wang, Dong and Cao, Hui and He, Wanli and Pan, Feng and others},
  journal={Advanced Functional Materials},
  volume={31},
  number={43},
  pages={2104605},
  year={2021},
  publisher={Wiley Online Library}
}

@article{dang2020bulk,
  title={Bulk chiral halide perovskite single crystals for active circular dichroism and circularly polarized luminescence},
  author={Dang, Yangyang and Liu, Xiaolong and Sun, Yajing and Song, Jiewu and Hu, Wenping and Tao, Xutang},
  journal={The journal of physical chemistry letters},
  volume={11},
  number={5},
  pages={1689--1696},
  year={2020},
  publisher={ACS Publications}
}

@article{wang2025structural,
  title={Structural Design and Applications of Chiral Perovskites},
  author={Wang, Ze and Zhao, Guanghan and Zhang, Hong and Zhou, Huiqiong and Tang, Zhiyong},
  journal={Energy Material Advances},
  volume={6},
  pages={0305},
  year={2025},
  publisher={AAAS}
}

@article{kwon2025ligand,
  title={Ligand-Driven Chirality in Perovskites for Advanced Optoelectronics},
  author={Kwon, Boesung and Park, Jonghyun and Choi, Wonbin and Song, Haeni and Oh, Joon Hak},
  journal={ACS Applied Materials \& Interfaces},
  volume={17},
  number={39},
  pages={54356--54379},
  year={2025},
  publisher={ACS Publications}
}

@article{makhija2024effect,
  title={Effect of film morphology on circular dichroism of low-dimensional chiral hybrid perovskites},
  author={Makhija, Urmila and Rajput, Parikshit Kumar and Parthiban, Pavithra and Nag, Angshuman},
  journal={The Journal of Chemical Physics},
  volume={160},
  number={2},
  year={2024},
  publisher={AIP Publishing}
}

@article{salij2021theory,
  title={Theory of apparent circular dichroism reveals the origin of inverted and noninverted chiroptical response under sample flipping},
  author={Salij, Andrew and Goldsmith, Randall H and Tempelaar, Roel},
  journal={Journal of the American Chemical Society},
  volume={143},
  number={51},
  pages={21519--21531},
  year={2021},
  publisher={ACS Publications}
}

@article{lv_self-assembled_2022,
	title = {Self-assembled inorganic chiral superstructures},
	volume = {6},
	issn = {2397-3358},
	url = {https://doi.org/10.1038/s41570-021-00350-w},
	doi = {10.1038/s41570-021-00350-w},
	number = {2},
	journal = {Nature Reviews Chemistry},
	author = {Lv, Jiawei and Gao, Xiaoqing and Han, Bing and Zhu, Yanfei and Hou, Ke and Tang, Zhiyong},
	month = {feb},
	year = {2022},
	pages = {125--145},
}

@article{ranjbar2009circular,
  title={Circular dichroism techniques: biomolecular and nanostructural analyses-a review},
  author={Ranjbar, Bijan and Gill, Pooria},
  journal={Chemical biology \& drug design},
  volume={74},
  number={2},
  pages={101--120},
  year={2009},
  publisher={Wiley Online Library}
}

@article{moss1996basic,
  title={Basic terminology of stereochemistry (IUPAC Recommendations 1996)},
  author={Moss, Gerry P},
  journal={Pure and applied chemistry},
  volume={68},
  number={12},
  pages={2193--2222},
  year={1996},
  publisher={De Gruyter}
}

@book{berova2012comprehensive,
  title        = {Comprehensive Chiroptical Spectroscopy, Volume 1: Instrumentation, Methodologies, and Theoretical Simulations},
  editor       = {Berova, Nina and Polavarapu, Prasad L and Nakanishi, Koji and Woody, Robert W},
  volume       = {1},
  year         = {2012},
  publisher    = {John Wiley \& Sons},
}

@article{soares2025chirality,
  title={Chirality-driven dimensionality and broadband emission in lead bromide perovskites},
  author={Soares, C{\'a}ssio CS and Acosta, Carlos Mera and Ferreira, F{\'a}bio F and Tofanello, Aryane and G{\'o}mez, Mayra AP and Ayala, Alejandro P and Toledo, JR and Gobato, Yara Galv{\~a}o and Lemes, Maykon A and Paschoal, Carlos WA and others},
  journal={Materials Advances},
  volume={6},
  number={24},
  pages={9545--9555},
  year={2025},
  publisher={Royal Society of Chemistry}
}

@article{kahmann2025pathways,
  title={Pathways, Probes, and Puzzles of Broadband Luminescence in “Perovskite-Inspired” Materials},
  author={Kahmann, Simon},
  journal={ACS Materials Letters},
  volume={7},
  number={5},
  pages={1732--1736},
  year={2025},
  publisher={ACS Publications}
}

@article{rodger2016linear,
  title={Linear dichroism as a probe of molecular structure and interactions},
  author={Rodger, Alison and Dorrington, Glen and Ang, Dale L},
  journal={Analyst},
  volume={141},
  number={24},
  pages={6490--6498},
  year={2016},
  publisher={Royal Society of Chemistry}
}

@article{ugras2023can,
  title={Can we still measure circular dichroism with circular dichroism spectrometers: The dangers of anisotropic artifacts},
  author={Ugras, Thomas J and Yao, Yuan and Robinson, Richard D},
  journal={Chirality},
  volume={35},
  number={11},
  pages={846--855},
  year={2023},
  publisher={Wiley Online Library}
}

@article{narushima2016circular,
  title={Circular dichroism microscopy free from commingling linear dichroism via discretely modulated circular polarization},
  author={Narushima, Tetsuya and Okamoto, Hiromi},
  journal={Scientific Reports},
  volume={6},
  number={1},
  pages={35731},
  year={2016},
  publisher={Nature Publishing Group UK London}
}

@book{shrivastava2018introduction,
  title={Introduction to plastics engineering},
  author={Shrivastava, Anshuman},
  year={2018},
  publisher={William Andrew}
}

@article{li2025spatial,
  title={Spatial mapping of chiral-induced spin selectivity in chiral perovskite via spin-Schottky junction},
  author={Li, Minghui and Chen, Zhongwei and Lang, Xiting and Zhang, Junchuan and Jiang, Yongjie and Tian, Hao and Ye, Fang and Liu, Xirui and Gou, Yangyang and Xi, Herui and others},
  journal={National Science Review},
  volume={12},
  number={9},
  pages={nwaf295},
  year={2025},
  publisher={Oxford University Press}
}

@article{chiesa2021assessing,
  title={Assessing the nature of chiral-induced spin selectivity by magnetic resonance},
  author={Chiesa, Alessandro and Chizzini, Mario and Garlatti, E and Salvadori, E and Tacchino, Francesco and Santini, Paolo and Tavernelli, Ivano and Bittl, Robert and Chiesa, M and Sessoli, R and others},
  journal={The Journal of Physical Chemistry Letters},
  volume={12},
  number={27},
  pages={6341--6347},
  year={2021},
  publisher={ACS Publications}
}

@article{wei2021chiral,
  title={Chiral perovskite spin-optoelectronics and spintronics: toward judicious design and application},
  author={Wei, Qi and Ning, Zhijun},
  journal={ACS Materials Letters},
  volume={3},
  number={9},
  pages={1266--1275},
  year={2021},
  publisher={ACS Publications}
}

@article{castiglioni2010evaluation,
  title={Evaluation of instrumental errors built in circular dichroism spectrometers},
  author={Castiglioni, Ettore and Albertini, Paolo and Abbate, Sergio},
  journal={Chirality},
  volume={22},
  number={1E},
  pages={E142--E148},
  year={2010},
  publisher={Wiley Online Library}
}

@article{kuroda2024fast,
  title={Fast and artifact-free circular dichroism measurement of solid-state structural changes},
  author={Kuroda, Reiko and Harada, Takunori and Takahashi, Hiromi},
  journal={Chirality},
  volume={36},
  number={1},
  pages={e3622},
  year={2024},
  publisher={Wiley Online Library}
}

@article{yang2017emergent,
  title={Emergent properties of an organic semiconductor driven by its molecular chirality},
  author={Yang, Ying and Rice, Beth and Shi, Xingyuan and Brandt, Jochen R and Correa da Costa, Rosenildo and Hedley, Gordon J and Smilgies, Detlef-M and Frost, Jarvist M and Samuel, Ifor DW and Otero-De-La-Roza, Alberto and others},
  journal={Acs Nano},
  volume={11},
  number={8},
  pages={8329--8338},
  year={2017},
  publisher={ACS Publications}
}

@book{Kobayashi2011CDOrgChem,
    author = {Kobayashi, Nagao and Muranaka, Atsuya and Mack, John},
    title = {Circular Dichroism and Magnetic Circular Dichroism Spectroscopy for Organic Chemists},
    publisher = {The Royal Society of Chemistry},
    year = {2011},
    month = {11},
    isbn = {978-1-84755-869-5},
    doi = {10.1039/9781849732932},
    url = {https://doi.org/10.1039/9781849732932},
}

@article{ferre1984linear,
  title={Linear optical birefringence of magnetic crystals},
  author={Ferr{\'e}, J and Gehring, GA},
  journal={Reports on Progress in Physics},
  volume={47},
  number={5},
  pages={513},
  year={1984},
  publisher={IOP Publishing}
}

@article{asad2024realization,
  title={Realization of giant superstructural chirality at broadband optical wavelengths via perovskite dielectric metasurfaces},
  author={Asad, Aqsa and Khaliq, Hafiz Saad and Kim, Min-Seok and Lee, Jae-Won and Kim, Hak-Rin},
  journal={Materials Advances},
  volume={5},
  number={6},
  pages={2536--2544},
  year={2024},
  publisher={Royal Society of Chemistry}
}

@article{guo2025tunable,
  title={Tunable chiroptical activity and second-harmonic generation in chiral one-dimensional perovskites},
  author={Guo, Zhihang and Qiu, Xin and He, Tingchao and Wang, Changshun},
  journal={Chemical Communications},
  volume={61},
  number={74},
  pages={14157--14160},
  year={2025},
  publisher={Royal Society of Chemistry}
}

@article{rosenfeld1929quantenmechanische,
  title={{Quantenmechanische} {Theorie} der nat{\"u}rlichen optischen {Aktivit{\"a}t} von {Fl{\"u}ssigkeiten} und {Gasen}},
  author={Rosenfeld, L},
  journal={Zeitschrift f{\"u}r Physik},
  volume={52},
  number={3},
  pages={161--174},
  year={1929},
  publisher={Springer}
}

@misc{EManiMwebsite,
  author       = {{EManiM Project}},
  title        = {EManiM: Electronic Manifold for Material Informatics and Metrology},
  year         = {2026},
  url          = {https://emanim.szialab.org/index.html},
  note         = {Accessed: Feb 13, 2026}
}

@article{albano2017chiroptical,
  title={Chiroptical response inversion upon sample flipping in thin films of a chiral benzo [1, 2-b: 4, 5-b′] dithiophene-based oligothiophene},
  author={Albano, Gianluigi and Lissia, Margherita and Pescitelli, Gennaro and Aronica, Laura Antonella and Di Bari, Lorenzo},
  journal={Materials Chemistry Frontiers},
  volume={1},
  number={10},
  pages={2047--2056},
  year={2017},
  publisher={Royal Society of Chemistry}
}

@article{ishii2020direct,
  title={Direct detection of circular polarized light in helical 1D perovskite-based photodiode},
  author={Ishii, A and Miyasaka, T},
  journal={Science advances},
  volume={6},
  number={46},
  pages={eabd3274},
  year={2020},
  publisher={American Association for the Advancement of Science}
}

@article{takechi2011chiroptical,
  title={Chiroptical measurement of chiral aggregates at liquid-liquid interface in centrifugal liquid membrane cell by Mueller matrix and conventional circular dichroism methods},
  author={Takechi, Hideaki and Arteaga, Oriol and Ribo, Josep M and Watarai, Hitoshi},
  journal={Molecules},
  volume={16},
  number={5},
  pages={3636--3647},
  year={2011},
  publisher={MDPI}
}

@article{moroni2024chiral,
  title={Chiral 2D and quasi-2D hybrid organic inorganic perovskites: from fundamentals to applications},
  author={Moroni, Marco and Coccia, Clarissa and Malavasi, Lorenzo},
  journal={Chemical Communications},
  volume={60},
  number={70},
  pages={9310--9327},
  year={2024},
  publisher={Royal Society of Chemistry}
}

@article{kim2024giant,
  title={Giant chiral amplification of chiral 2D perovskites via dynamic crystal reconstruction},
  author={Kim, Hongki and Choi, Wonbin and Kim, Yu Jin and Kim, Jihoon and Ahn, Jaeyong and Song, Inho and Kwak, Minjoon and Kim, Jongchan and Park, Jonghyun and Yoo, Dongwon and others},
  journal={Science Advances},
  volume={10},
  number={34},
  pages={eado5942},
  year={2024},
  publisher={American Association for the Advancement of Science}
}

@article{arwin2021optical,
  title={Optical chirality determined from Mueller matrices},
  author={Arwin, Hans and Schoeche, Stefan and Hilfiker, James and Hartveit, Mattias and J{\"a}rrendahl, Kenneth and Ju{\'a}rez-Rivera, Olga Rubi and Mendoza-Galv{\'a}n, Arturo and Magnusson, Roger},
  journal={Applied Sciences},
  volume={11},
  number={15},
  pages={6742},
  year={2021},
  publisher={MDPI}
}

\newpage
\end{document}